\documentclass[onecolumn,letterpaper,aps,prc,longbibliography,superscriptaddress,showpacs,nofootinbib,floatfix,notitlepage]{revtex4-1}
\usepackage{lipsum}
\usepackage{graphicx,bm}
\usepackage{subcaption}
\usepackage{dblfloatfix}

\graphicspath{{./figs/antikt/}}

\usepackage{enumerate}
\usepackage{floatrow}
\usepackage{amssymb}
\usepackage{amsmath}
\usepackage{commath}
\usepackage{verbatim}
\usepackage{xspace}
\usepackage{color}
\usepackage[%
  colorlinks=true,
  urlcolor=blue,
  linkcolor=blue,
  citecolor=blue
]{hyperref}

\usepackage{etoolbox}
\makeatletter
\patchcmd\linenumberpar{\@LN@parpgbrk}{\penalty\@LN@parpgpen\relax}{}{}
\makeatother

\include{defs}

\begin{document}
\title{Searching for the dead cone effects with iterative declustering of heavy-flavor jets}

\newcommand{\Linst}{Oak Ridge National Laboratory, Physics Division, Oak Ridge, 37831 TN, USA}
\newcommand{\lbl}{Lawrence Berkeley National Laboratory, 1 Cyclotron Rd, Berkeley, 94720 CA, USA}
\affiliation{\Linst}
\affiliation{\lbl}

\author{Leticia Cunqueiro} \affiliation{\Linst}
\author{Mateusz P\l osko\'n} \affiliation{\lbl}

\begin{abstract}
We present a new method to expose the dead cone effect at colliders using iterative declustering techniques. Iterative declustering allows to unwind the jet clustering and to access the subjets or branches at different depths of the jet tree.
 Our method consists of declustering the heavy flavour-tagged jet
using Cambridge-Achen algorithm following the branch containing the heavy flavour at each step and registering the kinematics of the complementary untagged prong. The kinematics of the complementary untagged
prong fill a Lund map representing the gluon radiation off the heavy flavour quark at each step of the vacuum shower.

Using Pythia8 MC, we show that a simple cut on the Lund plane introduced by ln($k_{\rm T}$) $> 0$, suppresses hadronisation effects and the angular separation between the jet prongs becomes very sensitive to flavour effects. A clear suppression for heavy flavour jets relative to inclusive jets in the region of splitting angles delimited by the relation $\theta < m_{Q}/E$ is observed, where $m_{Q}$ is the mass of the heavy quark and $E$ is the energy of the radiator or splitting prong.
\end{abstract}
\maketitle
\section{Introduction}
\label{S:1}

The dead cone effect is a fundamental prediction of QCD (and gauge theories in general) according to which the radiation from a charged particle of mass m and energy E is suppressed at angular scales given by m/E \cite{Dokshitzer:1991fd}.

To experimentally uncover the dead cone is a difficult task. The decays of the heavy flavor particles happen at similar angular scales and fill the deadcone.
The irresolution of the axis chosen as proxy for the heavy flavor direction can also obscure its measurement. In  \cite{Battaglia:2004coa} the first direct measurement in e$^{+}$e$^{-}$ collisions of the depletion of fragmentation particles not coming from the heavy flavor decay vertex in heavy-flavor tagged jets was presented. The main limiting element was the choice of the reference axis, which was estimated using the thrust direction, the jet axis and the vertex direction.

Recently new ideas were proposed to measure the dead cone using Soft Drop grooming techniques and boosted top quarks at the Large Hadron Collider \cite{Maltoni:2016ays}.

In this paper we discuss the use of new iterative declustering techniques \cite{Dreyer:2018nbf, Andrews:2018jcm} that allow to penetrate the jet shower and access the deepest levels of the clustering history which correspond to the splittings at the smallest angles. This may allow to unveil the dead cone even for low mass heavy flavors such as charm and beauty, provided they can be fully reconstructed in the experiment and that the angular resolution of the detector is good enough to separate subjets at distances of order 0.1 radians, which is typically the case with a tracking device.

\section{Iterative declustering of heavy flavor-tagged jets} \label{S:2}

We have generated Pythia 8.226 Tune 4C \cite{Sjostrand:2014zea} hard events leading to charm and
beauty quark-initiated jets and we have inhibited the decay of the D and
B mesons. We reconstruct the jets with anti-k$_{\rm T}$ algorithm \cite{Cacciari:2008gp} and
tag the jets when a D$^{0}$ or B$^{0}$ meson is found as one of its constituents.

The tagged jets are then declustered using the Cambridge/Aachen (C/A) algorithm \cite{Dokshitzer:1997in}. Since the
C/A metric is such that particles at small angles are combined first,
the first steps of the declustering process find the prongs at the largest
relative angles.

When moving backwards through the jet history, we always follow the branch containing the heavy flavor and we register the relative transverse
momentum $k_{\rm T}$ and angle $\theta$ of the complementary untaged prong onto the Lund map.
The cases where the heavy-flavor tagged prong is not the hardest are of the order of $1\%$ and $ 0.001\%$ in the explored kinematic regime for c and b-jets respectively and their exclusion from the Lund map has no quantitative impact on the results, so there are no ambiguities in the comparison to the inclusive jets, in which case we follow the hardest prong.

\begin{figure}
\begin{subfigure}{\linewidth}
\includegraphics[width=0.49\textwidth]{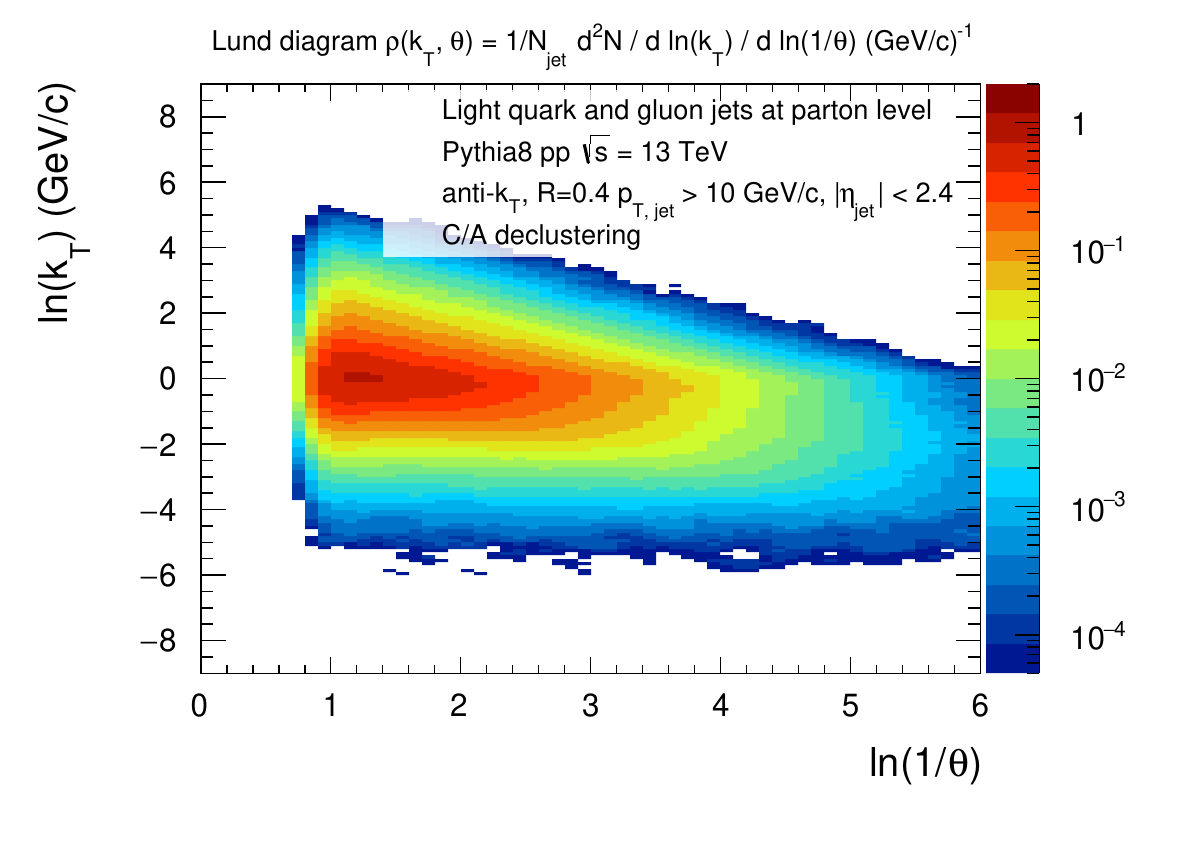}
\includegraphics[width=0.49\textwidth]{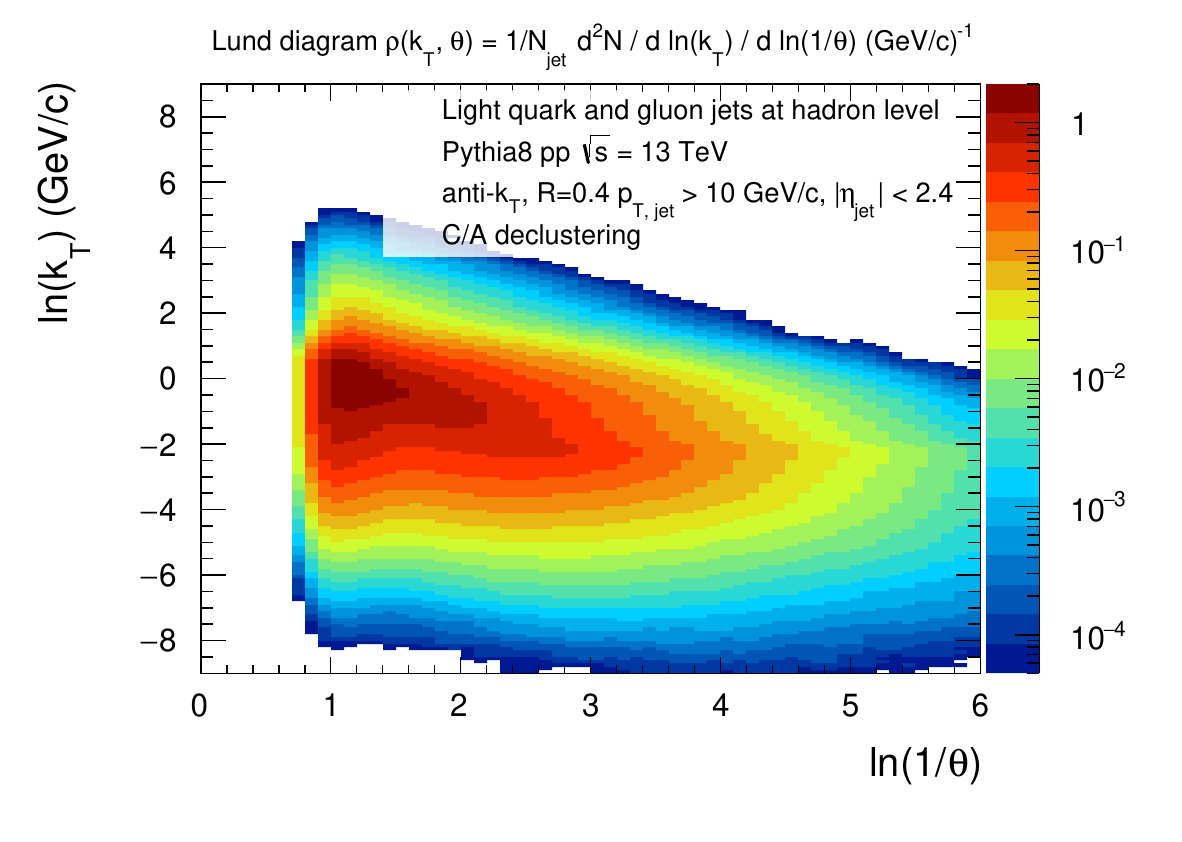}
\end{subfigure}
\begin{subfigure}{\linewidth}
\includegraphics[width=0.49\textwidth]{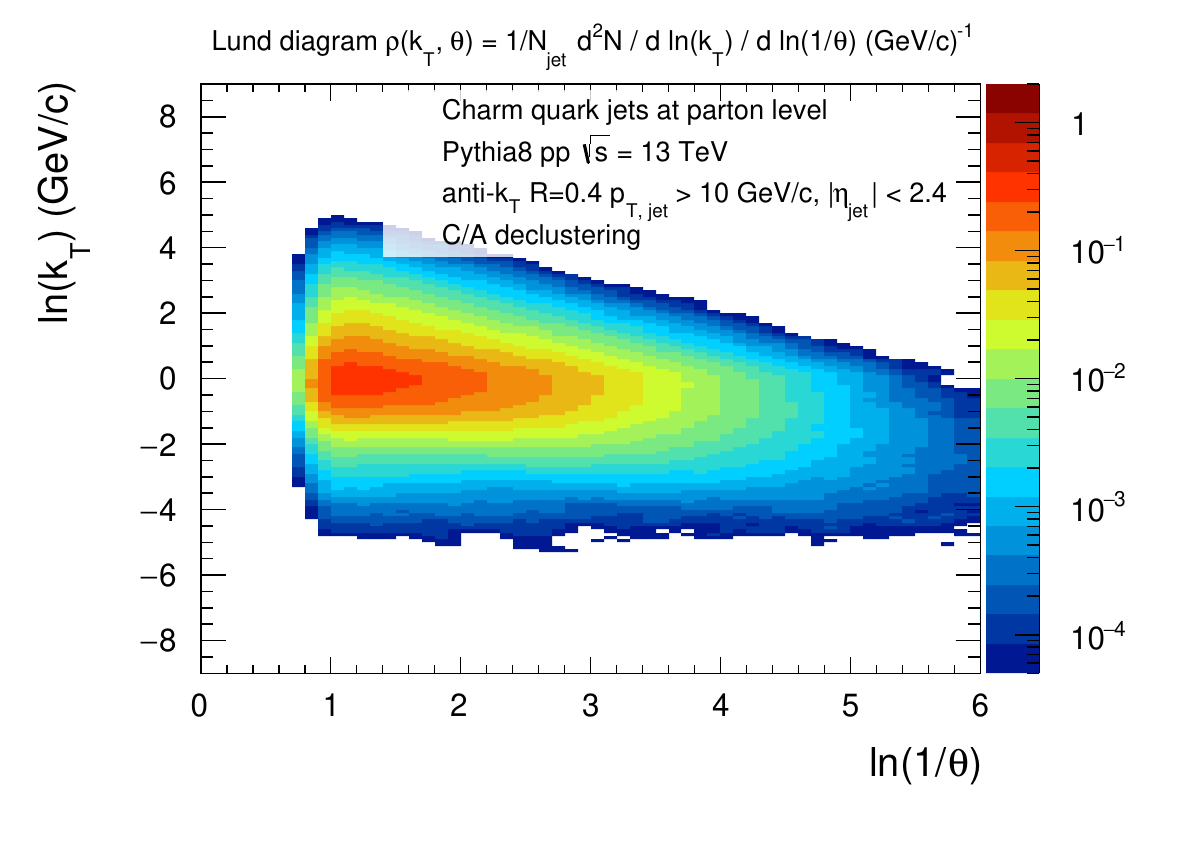}
\includegraphics[width=0.49\textwidth]{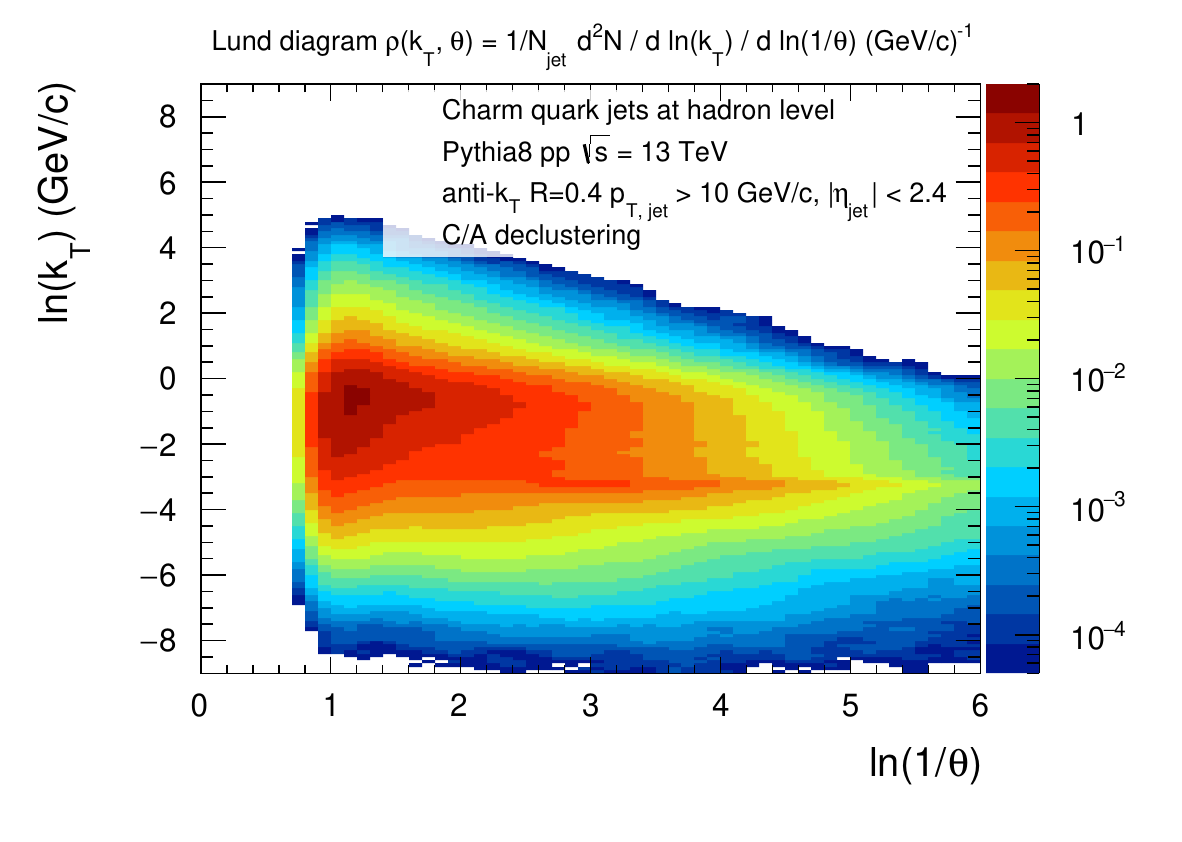}
\end{subfigure}
\begin{subfigure}{\linewidth}
\includegraphics[width=0.49\textwidth]{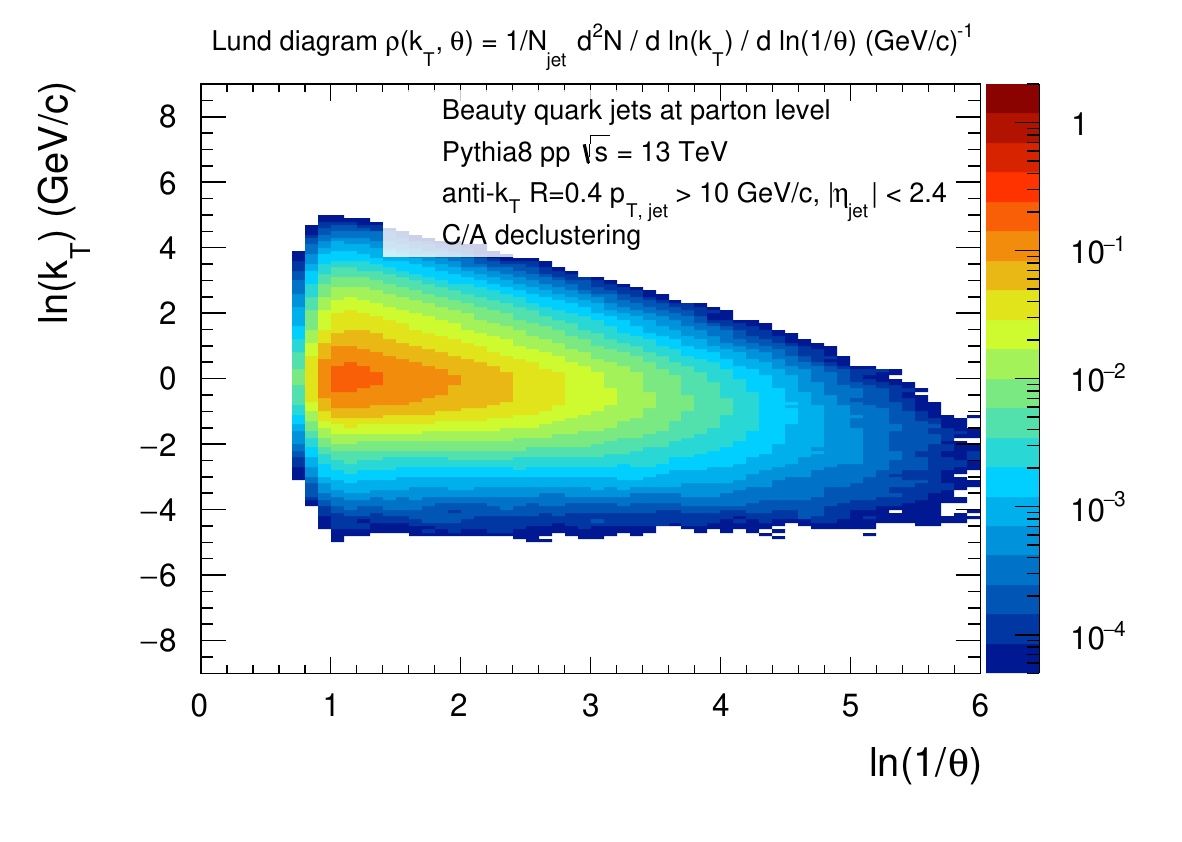}
\includegraphics[width=0.49\textwidth]{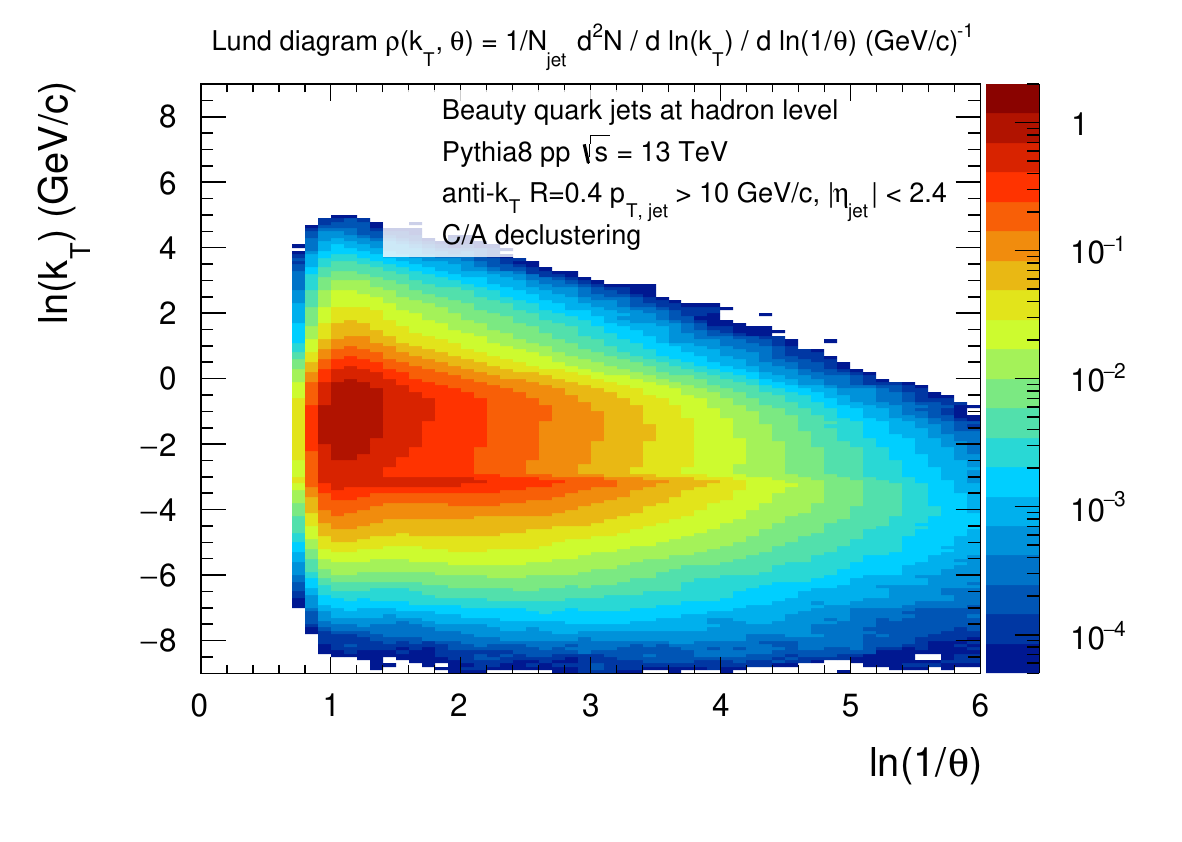}
\end{subfigure}
\caption{Left column: Lund diagrams for charm and beauty and inclusive jets at parton level and UE switched off, for low momentum jets of $10 < p_{T,jet} < 40$. Right column: Same plots but at hadron level. }
\label{fig:LundParton}
\end{figure}

Figure \ref{fig:LundParton} left column, shows the Lund maps for b and c-tagged jets
together with inclusive jets, at parton level. The $y$ and $x$ axes are ln($k_{\rm T}$) and ln($1/\theta$) respecively. The 2D maps are normalized to the total number of jets. The underlying event was switched off.

Already visually, without further analysis, one can note that the small angle region is less filled for heavy quarks jets than for inclusive jets.


Figure \ref{fig:LundParton} right column,  shows the Lund maps for b and c-tagged jets together with inclusive jets at hadron level. The hadronisation effects pollute the low $ln(k_{\rm T})$ part of the diagram, which corresponds to non-perturbative scales.
The UE pollutes the large-angle sector, where catchment area is maximal.
In order to suppress non-perturbative effects a simple cut on the splitting scale is possible: $ln(k_{\rm T})>0$, which selects splittings governed by scales of at least $1$ GeV/$c$.

\section{The dead cone}

\begin{figure}[h]
\begin{subfigure}{\linewidth}
\includegraphics[width=0.49\textwidth]{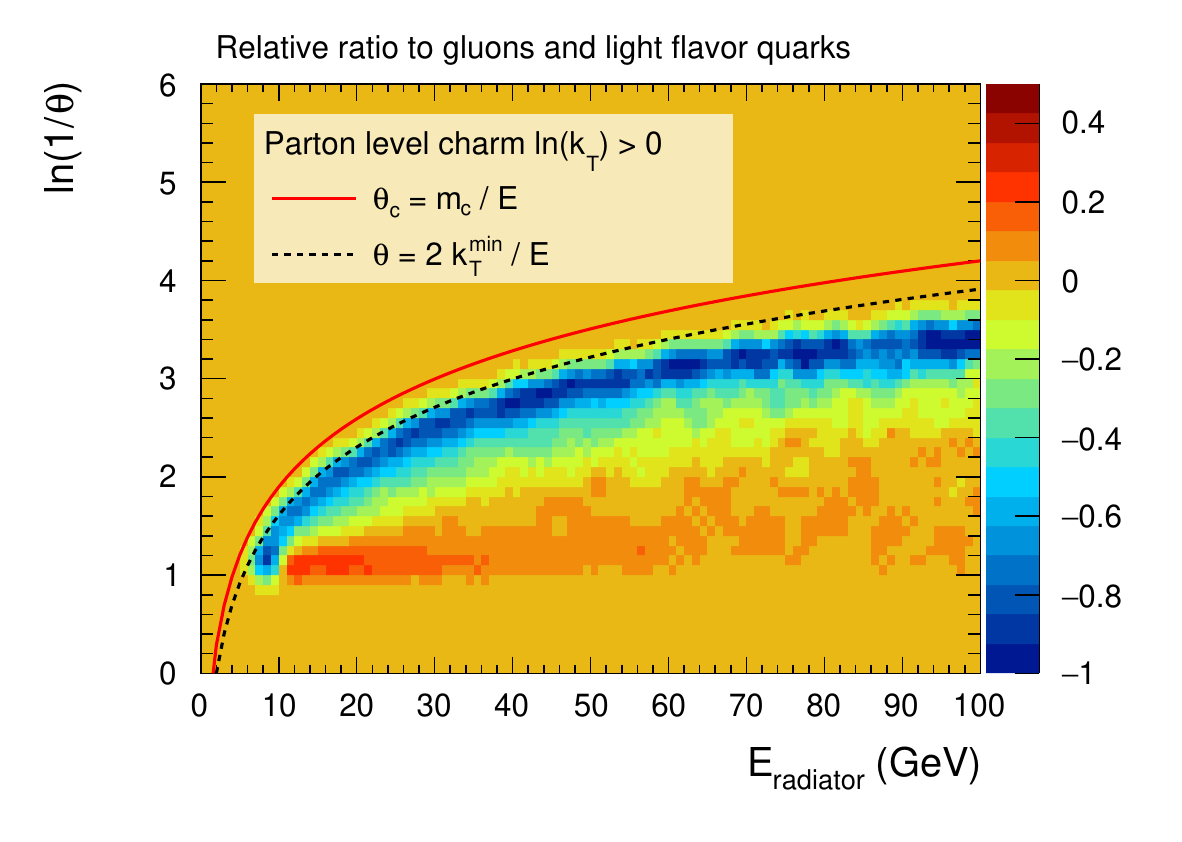}
\includegraphics[width=0.49\textwidth]{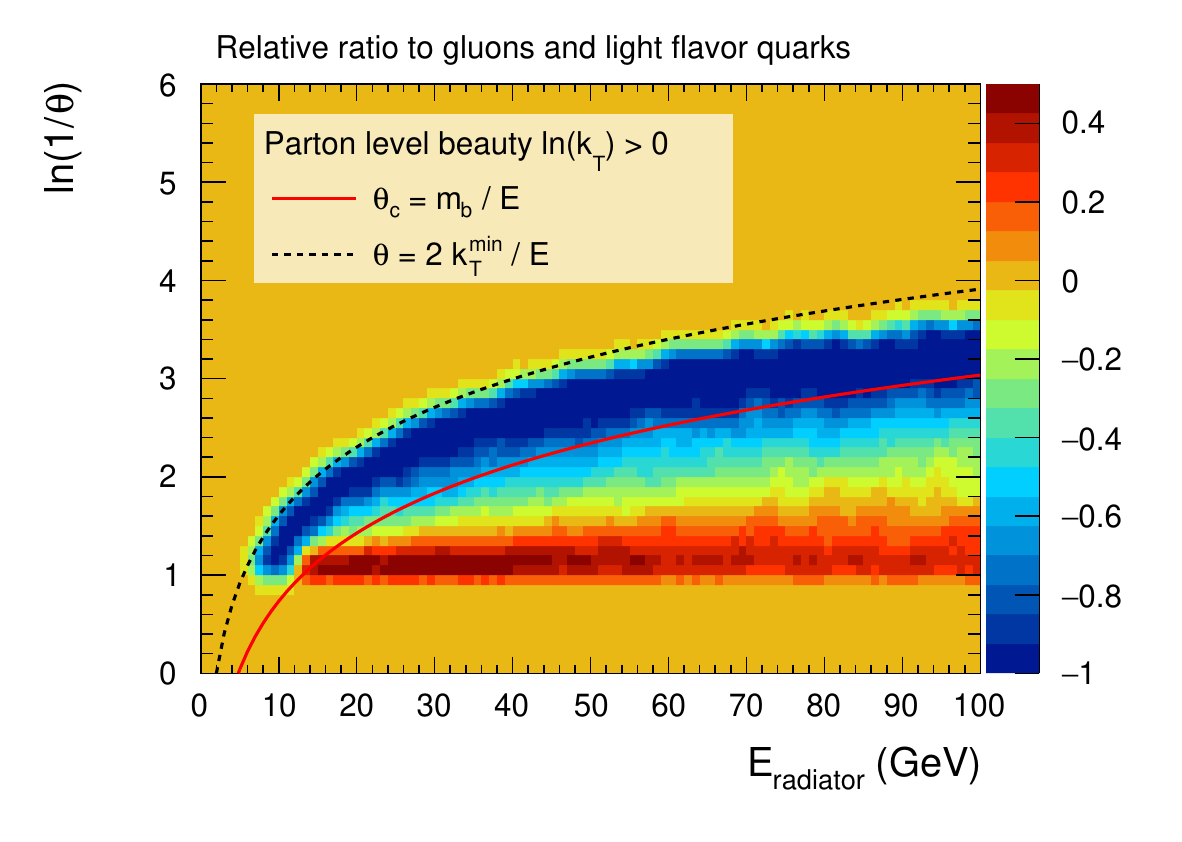}
\caption{Parton level c-jets (left) and b-jets (right).}
\label{fig:DeadConeParton}
\end{subfigure}
\begin{subfigure}{\linewidth}
\includegraphics[width=0.49\textwidth]{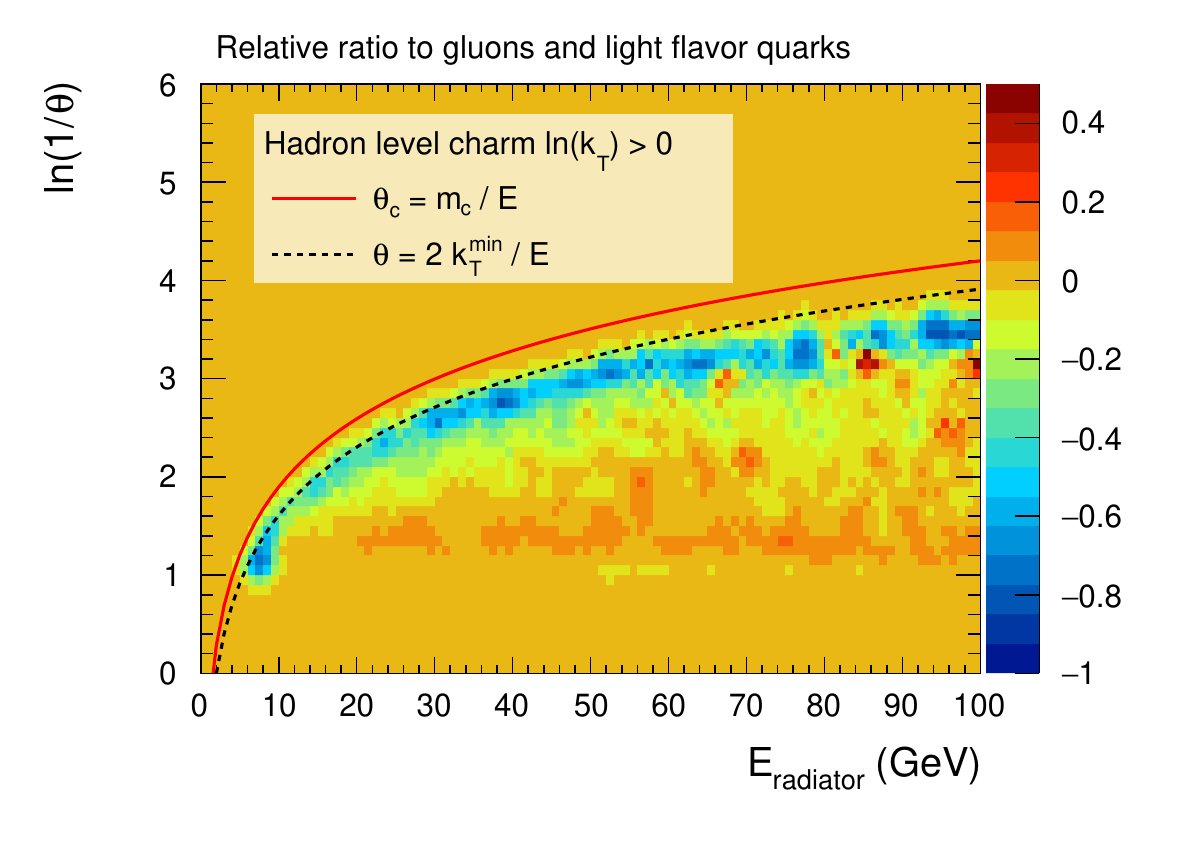}
\includegraphics[width=0.49\textwidth]{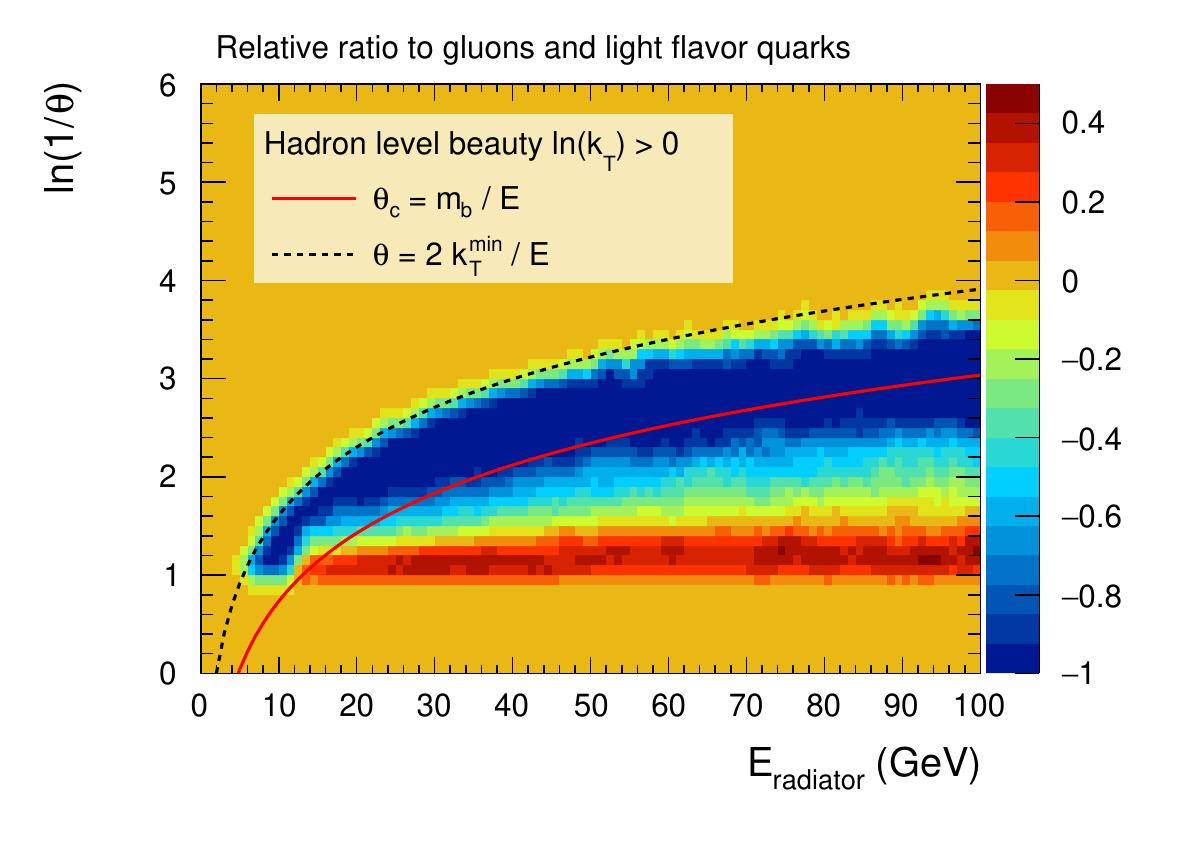}
\caption{Hadron level c-jets (left) and b-jets (right).}
\label{fig:DeadConeHadronFull}
\end{subfigure}
\begin{subfigure}{\linewidth}
\includegraphics[width=0.49\textwidth]{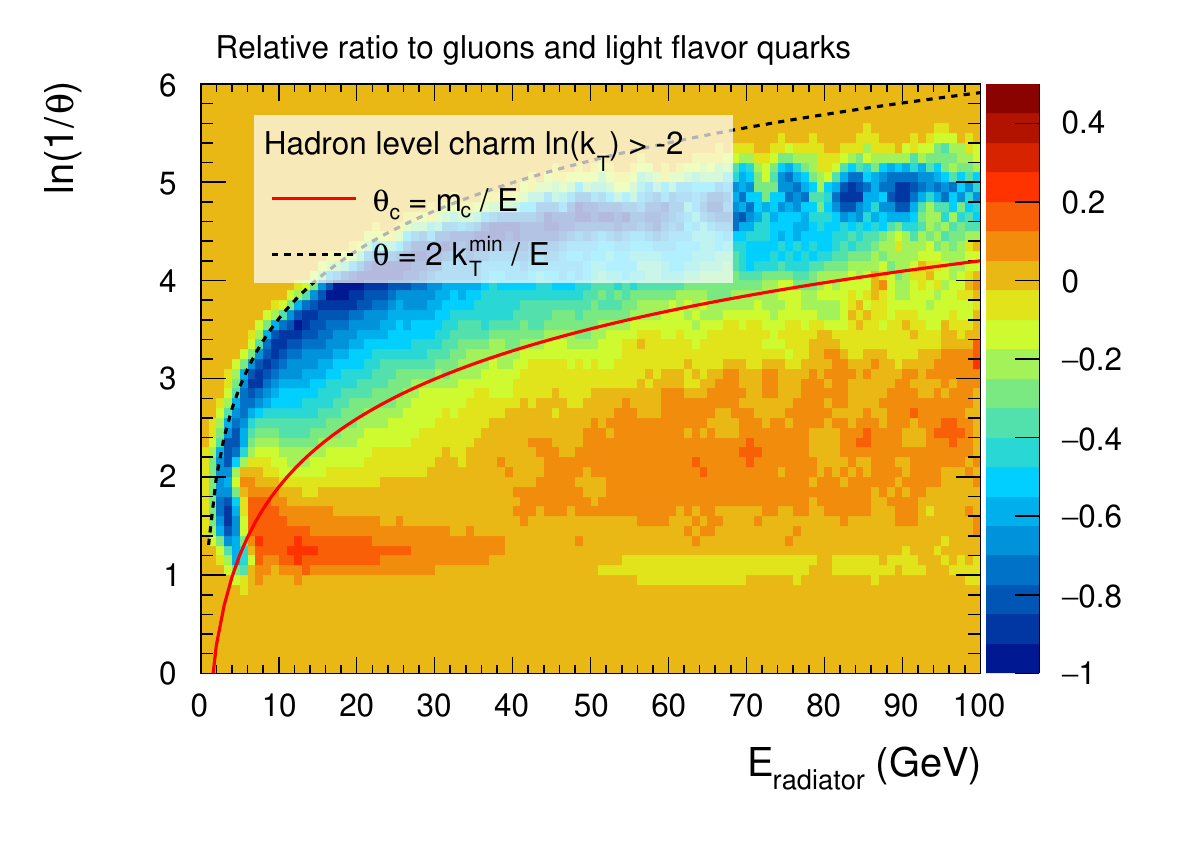}
\includegraphics[width=0.49\textwidth]{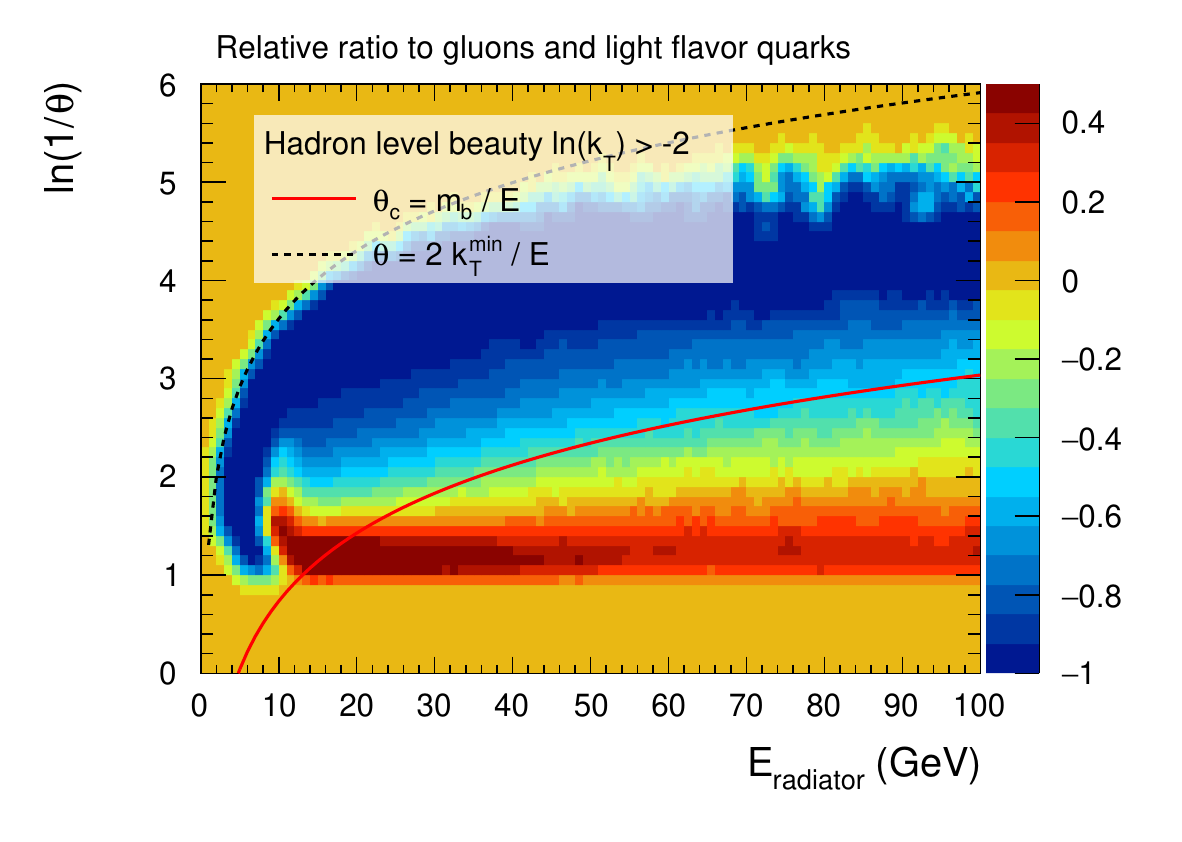}
\caption{Hadron level with a relaxed cut on $k_{\rm T}$ - demonstration of the impact of non-perturbative effects }
\label{fig:DeadconeDirty}
\end{subfigure}
\caption{Relative difference for heavy flavor and inclusive jets of the correlation of the splitting angle and the energy of the radiator. The red curves correspond to the critical angle $\theta_{C}=m_{Q}/E$. }
\label{fig:anglevsE}
\end{figure}

A different representation of the Lund diagram which exposes the dead cone effect more clearly is that where the horizontal axis corresponds to the energy of the radiating daughter prong and the vertical axis corresponds to the splitting angle $\theta$. We construct this 2D map and we consider the relative difference between heavy quark-tagged jets and inclusive jets as a way to study the relative enhancement/suppression in the different areas of interest:
\begin{equation}
	Q=\frac{P^{Q}(ln(1/\theta),E_{radiator})-P^{inc}(ln(1/\theta),E_{radiator})}{P^{inc}(ln(1/\theta),E_{radiator})}
\end{equation}
where $P(ln(1/\theta),E_{radiator})$ represents the probability for a a radiator prong with energy $E_{radiator}$ to split with an apperture angle $\theta$.
This is illustrated in Figure \ref{fig:DeadConeParton}. The cut $ln(k_{\rm T})>0$ translates into $E_{radiator}>1/(z \theta)$, where $z$ is the energy fraction carried by the daughter prong. The kinematic limit $z=0.5$  imposes the sharp threshold in the curve above which there are no entries in the inclusive reference.

One can clearly see a region of the phase space where $Q$ becomes negative, indicating a supression of the splittings for the heavy quarks compared to inclusive jets, both for c (left) and b jets (right). The angular suppression is coupled to a suppression of the high $z$ splittings: the suppression of small angles affects hard emissions with small $k_{\rm T}$, which are the most probable. This leads to a reduction of the intra-jet multiplicity and to a hardening of the fraction of energy kept by the leading prong in the jet.

The region of angles smaller than $\theta<\theta_{C}=m_{Q}/E$, corresponds to the area above the red curve. We note that the parametrical dead cone red line qualitatively coincides with the curve where $Q_{\rm{Beauty}}$ becomes $-1$, when radiation of the heavy flavor prong is completely suppressed, though the relation is not exact.
In the case of charm, the dead cone line is above the kinematic threshold for the given $k_{\rm T}$ selection. More agressive cuts on the scale $k_{\rm T}$ select more large angle radiation and reduce the range of observation of the dead cone related suppression.

At hadron level the effects are not washed out. A suppression of significant magnitude at angles smaller than the critical angle at all radiator energies for b and c jets is observed, see Figure \ref{fig:DeadConeHadronFull}.

\begin{figure}[tb]
\centering
\includegraphics[width=0.45\textwidth]{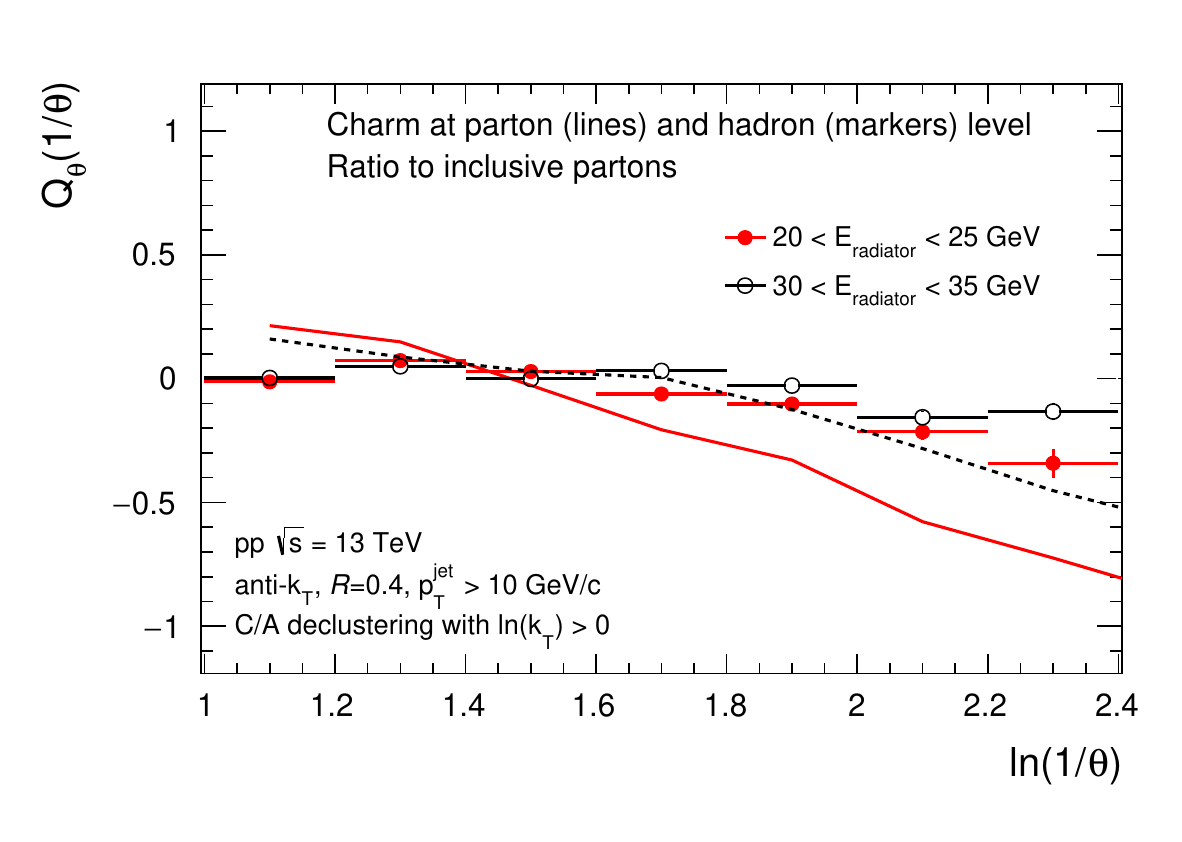}
\includegraphics[width=0.45\textwidth]{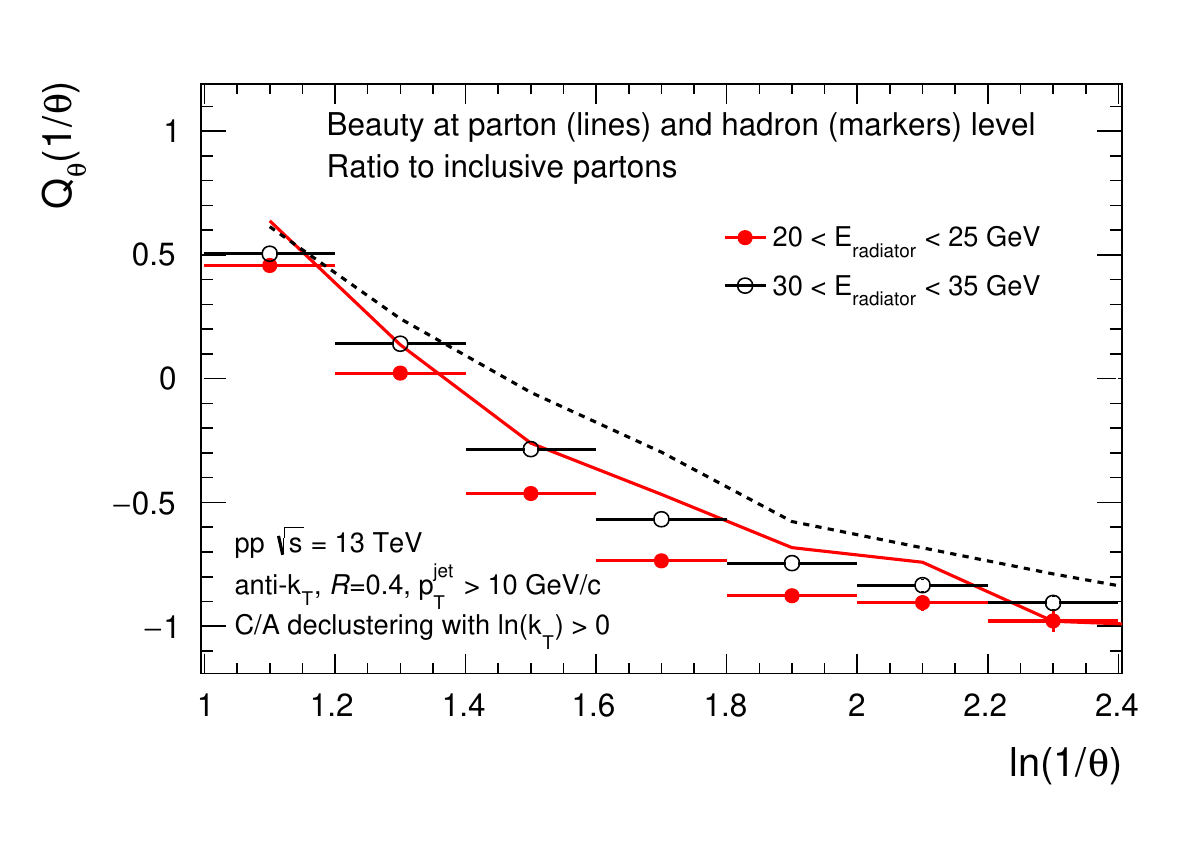}
\caption{Relative difference of the angular distribution for heavy flavor and inclusive jets, for perturbative scales defined by ln($k_{\rm T}>0$) and for different bins of radiator energy. Left and right plots correspond to Charm and Beauty jets while solid/dashed lines correspond to hadron/parton level. }
\label{fig:AngleDistrib}
\end{figure}

In Figure \ref{fig:DeadconeDirty} we show the impact of considering a looser cut ln($k_{\rm T} >-2$). Non-perturbative effects fill the dead cone and obscure the effect of the suppression of the radiation.
In Figure \ref{fig:AngleDistrib} we show our proposed observable, which is the projection onto the vertical axis of the diagrams shown in Fig. \ref{fig:DeadConeHadronFull}, for a range of low radiator energies, for instance  $20<E_{radiator}<50$ GeV. The observable, denoted as $Q_{\theta}$, corresponds to the relative difference between the angular distribution of the splittings for heavy flavor jets and inclusive jets, at perturbative scales set by ln($k_{\rm T})>0$ and for low radiator energies for which the dead cone effects are maximal:
\begin{equation}
	Q_{\theta}=\frac{P^{Q}(1/\theta)-P^{inc}(1/\theta)}{P^{inc}(1/\theta)}, E_{radiator} \in (E_{min},E_{max})
\end{equation}
The suppression of the low angle emission probability for b-tagged radiators relative to
inclusive ones is of order $80\%$ at $ln(1/\theta)=2$, which approximately corresponds 0.14 radians.  The corresponding suppression for c-tagged radiators is of order 20$\%$.

The ideal experiment would be able to fully reconstruct the heavy flavor hadrons and to tag prongs of very low momentum, and to  separate subjets at angular scales of 0.1 and below.

\begin{figure}[tb]
\centering
\includegraphics[width=0.45\textwidth]{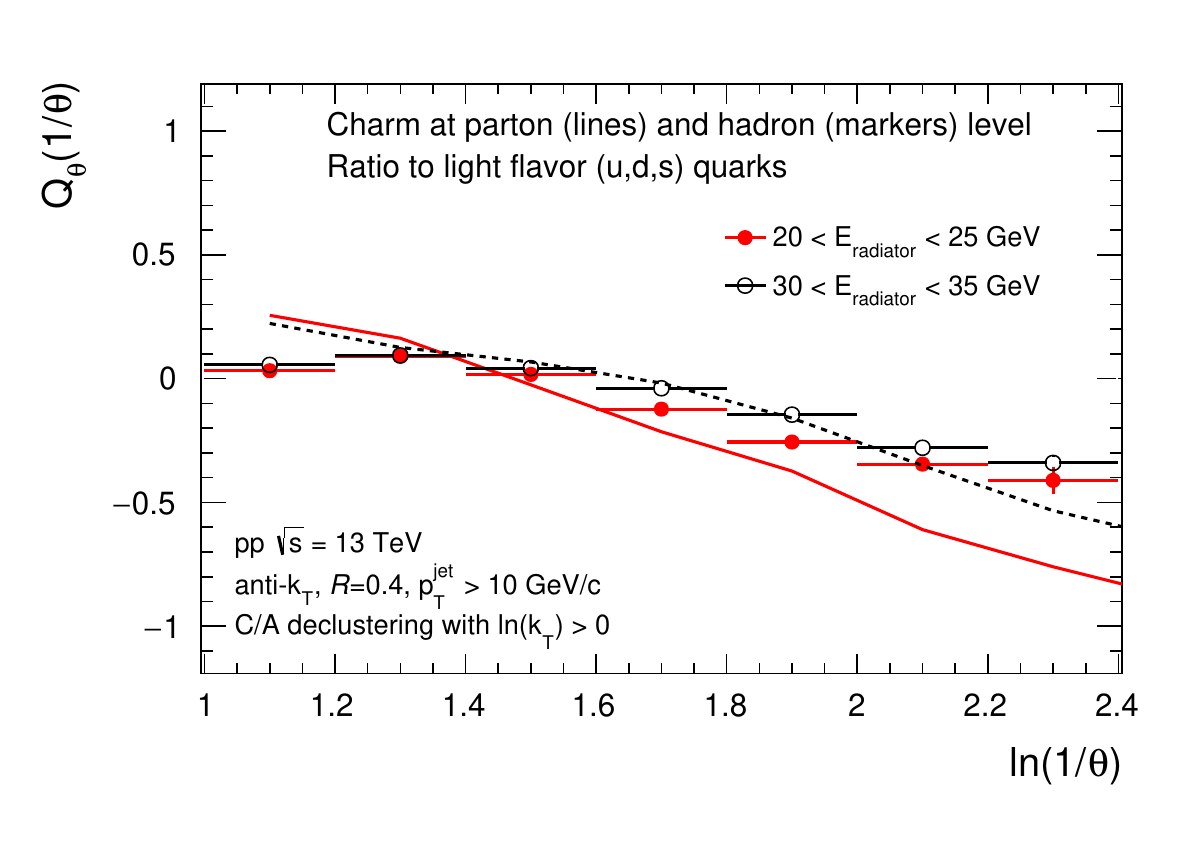}
\includegraphics[width=0.45\textwidth]{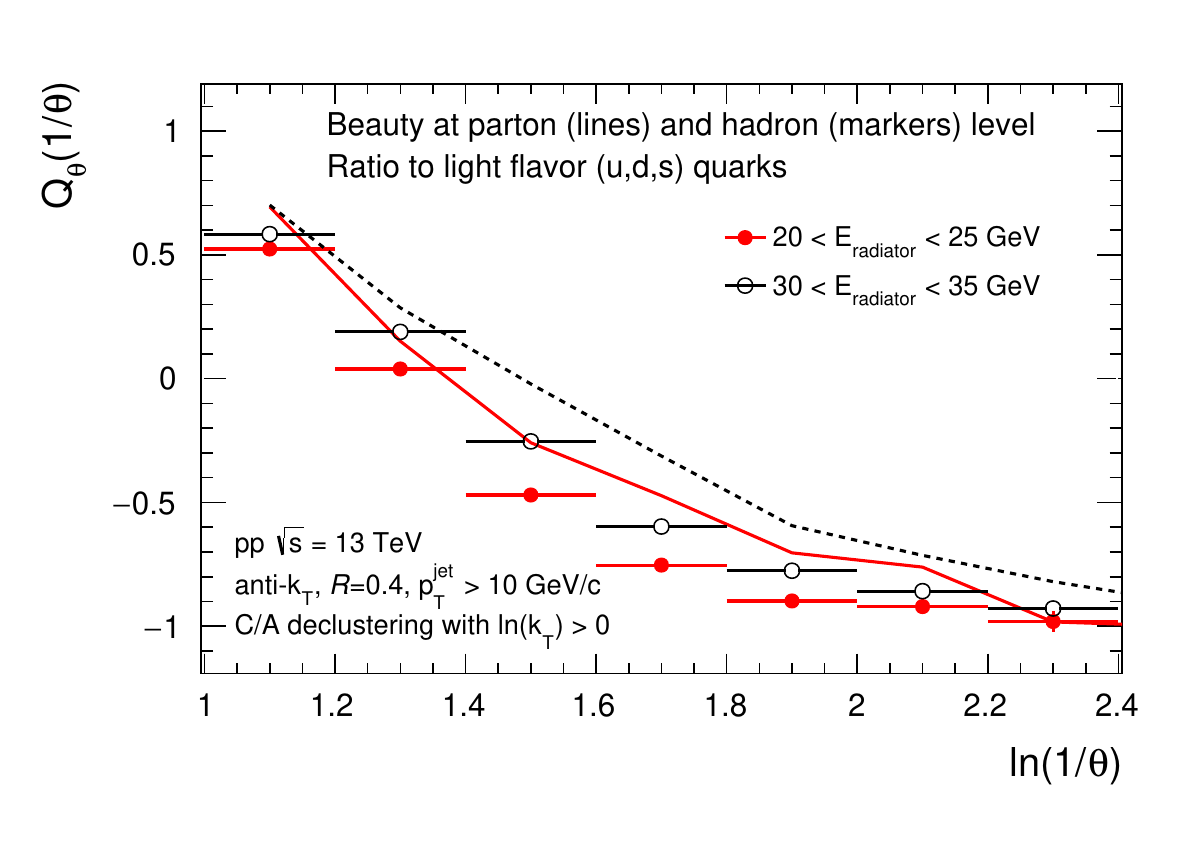}
\caption{Relative difference of the angular distribution for heavy flavor and light quark jets, for perturbative scales defined by ln($k_{\rm T}>0$) and for different bins of radiator energy. Left and right plots correspond to Charm and Beauty jets while solid/dashed lines correspond to hadron/parton level. }
\label{fig:AngleDistribQuarks}
\end{figure}

We note that the inclusive jets are not the best reference since they are a mixture of  $g \rightarrow gg$ and $q \rightarrow qg$ radiation with different fractions. The $g \rightarrow gg$ fill the phase space in a different way compared to $q \rightarrow qg$. We have checked that the magnitude of the suppression of $Q_{\theta}$ increases when light quark jets are considered as a reference instead of inclusive jets, see Fig. \ref{fig:AngleDistribQuarks}.
Experimentally it is possible to enrich the quark fraction by using boson-jet correlations for instance. Or by statistically cutting on observables that are sensitivie to differences between quark and gluon fragmentation such as the $p_{T}D$ or the jet angularity.

We have also tested that switching on/off the gluon splitting kernel $g \rightarrow qqbar$ has no impact on the results.

As a final remark we note that our method was tested only against Pythia8 and thus the exact magnitude of the dead-cone related effects and their onset rely on the specific Pythia8 implementation, which is done via matrix element corrections. However, parton shower generators feature the dead cone effect quite universally with better than $10\%$ agreement wit NLO calcuations \cite{Maltoni:2016ays}.

\section{Conclusions}

We have shown that iterative declustering is a tool that allows to access the deepest branches of the C/A jet trees which correspond to the smallest splitting angles and that can uncover quark mass  differences related to the dead cone.
We propose to build the Lund map for jets tagged with a fully reconstructed heavy flavor hadron. Furthermore, we propose to expose the perturbative splittings within the shower with a selection of $ln(k_{\rm T})>0$ and study the angular effects due to the deadcone at low radiator energies integrating over all declutering steps.  At angles of  order $0.1$ radians, the Pythia8 simulations predict a suppression of approximately 20/80$\%$ for jets containing a fully reconstructed D/B meson relative to inclusive jets, for radiator energies above 20 GeV.

\section*{Acknowledgements}
The authors thank Gavin Salam and Marco van Leeuwen for usefull discussions and comments to the manuscript.
This work was supported by the U.S. Department of Energy, Office of Science, Office of Nuclear Physics, under contracts DE-AC05-00OR22725 (ORNL) and DE-AC02-05CH11231 (LBNL).

\end{document}